\documentclass[aps,pre,twocolumn,eqsecnum,superscriptaddress]{revtex4}    
\usepackage{amsmath}    
\usepackage{amssymb}
\usepackage{subfigure}
\usepackage{epsfig}
\usepackage{array}      

\begin{document}

\title{Stick-slip instabilities in sheared granular flow: the role of friction and acoustic vibrations}

\author{Charles K. C. Lieou}
\affiliation{Department of Physics, University of California, Santa Barbara, CA 93106, USA}
\author{Ahmed E. Elbanna}
\affiliation{Department of Civil and Environmental Engineering, University of Illinois, Urbana-Champaign, IL 61801, USA}
\author{J. S. Langer}
\affiliation{Department of Physics, University of California, Santa Barbara, CA 93106, USA}
\author{J. M. Carlson}
\affiliation{Department of Physics, University of California, Santa Barbara, CA 93106, USA}
\date{\today}

\begin{abstract}
We propose a theory of shear flow in dense granular materials. A key ingredient of the theory is an effective temperature that determines how the material responds to external driving forces such as shear stresses and vibrations. We show that, within our model, friction between grains produces stick-slip behavior at intermediate shear rates, even if the material is rate-strengthening at larger rates. In addition, externally generated acoustic vibrations alter the stick-slip amplitude, or suppress stick-slip altogether, depending on the pressure and shear rate. We construct a phase diagram that indicates the parameter regimes for which stick-slip occurs in the presence and absence of acoustic vibrations of a fixed amplitude and frequency. These results connect the microscopic physics to macroscopic dynamics, and thus produce useful information about a variety of granular phenomena including rupture and slip along earthquake faults, the remote triggering of instabilities, and the control of friction in material processing.
\end{abstract}

\maketitle

\section{Introduction}

Stick-slip instabilities are ubiquitous in sheared granular materials ranging from pharmaceutical powders to earthquake fault gouge. Such instabilities are often observed in laboratory experiments \cite{johnson_2008,anthony_2005,mair_2002,han_2011,yamashita_2014,sone_2009,daniels_2014,
hayman_2011,marone_private}. At much larger scales, these instabilities may explain how aftershocks can be triggered by acoustic waves from earthquakes elsewhere~\cite{ferdowsi_2014a,ferdowsi_2014b,vanderelst_2010,griffa_2013,
daniels_2005,daniels_2006}. In the context of industrial processing, stick-slip could result in structural damage and problems in mixing, causing major problems in the quality control of products~\cite{merrow_2000,leuenberger_2005,roberts_2002}. However, only very few of the many grain-scale numerical simulations to date, such as~\cite{griffa_2011,daub_2011,mair_2008}, exhibit stick-slip behavior. Most other models do not account for the extended elastic interaction between the slipping region and its surroundings. It seems that we do not yet know what basic ingredients of granular models are necessary for the apparent rate-weakening behavior that produces the observed unstable behaviors. For both practical and theoretical purposes, we need a predictive, first-principles description of these phenomena.

In this paper we propose a nonequilibrium theory of stick-slip instabilities in dense granular flows. Our theory accounts for the effect of friction between colliding particles, and for the effects of both internally and externally driven acoustic vibrations. Our key findings are that interparticle friction is crucial for stick-slip instabilities, and that vibrations can amplify or reduce the stick-slip amplitude, or suppress stick-slip altogether, in different parameter regimes. A key ingredient of our analysis is the idea that the nonequilibrium states of a driven granular material are characterized by its compactivity
\begin{equation}\label{eq:X_def}
 X = \left( \dfrac{\partial V}{\partial S_C} \right)_{\{ \Lambda_{\alpha} \} },
\end{equation}
or equivalently, its effective disorder temperature $T_{\text{eff}} = p X$~\cite{haxton_2012,lieou_2012,lieou_2014a,lieou_2014b, edwards_1989a,edwards_1989b,edwards_1989c,edwards_1990a,edwards_1990b}. Here, $V$ is the extensive volume of the system, $S_C$ is the configurational entropy pertaining to slow, configurational degrees of freedom alone, and the $\Lambda_{\alpha}$'s are the internal variables, such as the density of defects and misalignments, that specify the configurational state of the granular material. The compactivity $X$ is controlled by how the system is driven, and in turn controls the volume. In fact, $X$ increases monotonically with the volume $V$ (if the grain mass is held constant), and decreases with increasing packing fraction. Increasing the shear rate would tend to dilate the system, while if one starts from a somewhat loose granular packing, increasing the vibration intensity would reduce the volume. This effective-temperature description is made possible by the fact that the slow, configurational degrees of freedom of the granular medium are driven out of equilibrium with the fast, kinetic-vibrational motions, so that the two sets of degrees of freedom are only weakly coupled to each other.

The basic idea is that the coupling ${\cal K}$ between the configurational and kinetic-vibrational subsystems consists of additive contributions of the fluctuations $\Gamma$ induced by shearing, $\xi$ induced by interparticle friction, and $\rho$ induced by external vibrations (when switched on). In our analysis~\cite{lieou_2014b} of a series of experiments by van der Elst \textit{et al}.~\cite{vanderelst_2012}, these are referred to as mechanical, frictional, and vibrational noise strengths. The effect of $\xi$ and $\rho$ on granular rheology is clearly illustrated there.

In the van der Elst \textit{et al}.~experiments, the authors measured the steady-state shear band thickness, or volume, as a function of the shear rate, for both spherical glass beads and angular sand particles in a cylindrical shear cell. The circular plate on top of the shear cell was rotated so that the shear band formed immediately underneath the plate, and contained roughly a dozen particles across its width. The apparatus was connected to a transducer which, when turned on, vibrated the grains at a single frequency and amplitude. Van der Elst \textit{et al.}~found that for unvibrated angular sand, the steady-state shear band thickness decreases as a function of increasing shear rate, at intermediate shear rates, but increases with shear rate in the fast, inertial flow regime. This is in contrast to smooth spherical glass beads for which the shear zone thickness increases monotonically with the shear rate. Moreover they found that external acoustic vibrations cause compaction in both angular sand and smooth glass beads at slow shear rates. Their experimental measurements and our quantitative fit to their data are shown in Fig.~1 in our earlier analysis \cite{lieou_2014b}. In Fig.~\ref{fig:xvplot} below, we also plot the dimensionless compactivity, instead of the volume or shear band thickness, as a function of the shear rate, albeit at a higher confining pressure than in the van der Elst \textit{et al}.~experiments.

Our analysis has shown that acoustic vibrations fluidize the granular material; that is, vibrations unjam the granular material and significantly reduce the flow stress at slow shear rates. In addition, frictional dissipation between particles suppresses shear dilatancy at intermediate shear rates, and is responsible for the nonmonotonic variation of the steady-state shear band thickness with shear rate seen in those experiments. This is shown by a direction comparison between the dashed and solid curves for frictionless and frictional particles in Fig.~\ref{fig:xvplot}. Nonmonotonic rheology is suggestive of inherent instabilities~\cite{dijksman_2011}, stick-slip failure being one example. This motivates our present theoretical analysis. We show in this paper that interparticle friction is indeed responsible for stick-slip at intermediate shear rates and high enough pressure. In addition, vibrations can amplify or reduce the stick-slip amplitude, or suppress stick-slip in different parameter regimes.

\begin{figure}[ht]
\centering 
\includegraphics[width=.45\textwidth]{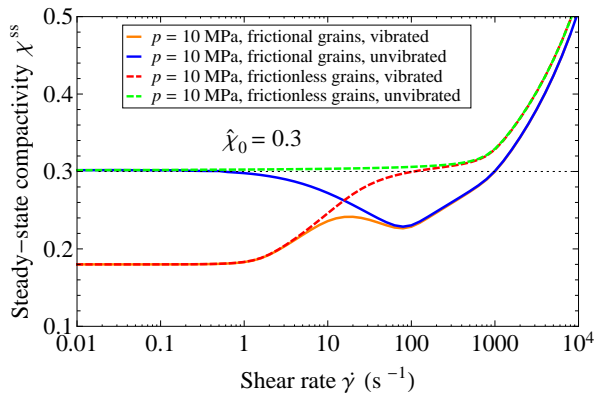}
\caption{\label{fig:xvplot}(Color online) A key ingredient for the emergence of flow instabilities is the nonmonotonic variation of the steady-state compactivity $\chi^{\text{ss}}$, as a function of the imposed shear rate $\dot{\gamma}$. (The quantity $\chi^{\text{ss}}$ increases monotonically with the layer thickness or volume, so the layer thickness follows the same qualitative behavior.) For clarity, this is shown here for a pressure $p = 10$ MPa, which lies below the critical pressure necessary for stick-slip to occur (see Fig.~\ref{fig:phase} below). It is implicitly assumed that physical insight gained from laboratory experiments at lower pressures can be extrapolated to higher pressures in the MPa range.} 
\end{figure}

The rest of the paper is structured as follows. Section~\ref{sec:2} is an exposition of the thermodynamic theory of driven granular media. There, we demonstrate how the first law of thermodynamics allows us to infer the manner in which the compactivity $X$ evolves as a function of time in sheared and/or vibrated granular media. This is followed by Sec.~\ref{sec:3} in which we decouple the effect of different driving mechanisms, and specify the configurational state of the granular media when either shearing and vibrations but not both is present. Then, in Sec.~\ref{sec:4}, we discuss the microscopic model of shear transformation zones (STZ's)~\cite{lieou_2012,falk_1998,langer_2011,lieou_2014a} which provide a direct connection between mechanical stress and irreversible, nonaffine granular rearrangement that accounts for dense granular flow. Upon completion of the theoretical development, we show in Sec.~\ref{sec:5} the emergence of stick-slip from our model, examine the effects of external vibrations at fixed frequency and amplitude, and discuss the conditions under which stick-slip instabilities are amplified or suppressed. We conclude our paper in Sec.~\ref{sec:6} with a discussion of open questions and future directions.

\section{Nonequilibrium thermodynamics of driven granular media}\label{sec:2}

The following theoretical development is almost the same as, but not identical to, that presented in Ref.~\cite{lieou_2014b}. Here, we explicitly account for the work done by external acoustic vibrations -- which need not be negligible -- and clarify the nature of the thermal temperature $T$. We also discuss how external vibrations directly control the configurational degrees of freedom of the system in Sec.~\ref{sec:3}.

Consider a granular medium in contact with a heat bath at temperature $T \approx 0$ (we set $T = 0$ later but retain the variable $T$ at present), with corresponding entropy $S_T$. Then, when the system is driven by a shear stress $s$ in the presence of a pressure $p$, and if vibrations do work on the medium at a rate $p\, V\, Y/ \tau$ (where the time scale $\tau$ is conveniently chosen to be the inertial time scale, discussed in Sec.~\ref{sec:4} below, and we have factored out $p \, V$ to make $Y$ dimensionless), the first law of thermodynamics reads
\begin{eqnarray}
\label{eq:first_law}
 \nonumber T \dot{S}_T &=&  V s\, \dot{\gamma}^{\text{pl}} - p\, \dot{V} + \dfrac{p\, V\, Y}{\tau} \\ &=&  V s\, \dot{\gamma}^{\text{pl}} - p X \dot{S}_C - p \sum_{\alpha} \left( \dfrac{\partial V}{\partial \Lambda_{\alpha}} \right)_{S_C} \dot{\Lambda}_{\alpha} + \dfrac{p\, V\, Y}{\tau}, ~~~~~
\end{eqnarray}
where $\dot{\gamma}^{\text{pl}}$ is the plastic shear rate. As in Eq.~\eqref{eq:X_def}, the $\Lambda_{\alpha}$'s are the internal variables, such as the density of flow defects, that specify the configuration of the granular medium. Note that while $T \approx 0$, the quantity $T \dot{S}_T$ need not be small, as energy is being dissipated into the environment. On the other hand, the second law of thermodynamics says that the rate of change of total entropy, being the sum of the thermal and configurational contributions, must be nonnegative:
\begin{equation}\label{eq:second_law}
 \dot{S} = \dot{S}_C + \dot{S}_T \geq 0 .
\end{equation}
Rearranging Eq.~\eqref{eq:first_law} to obtain an expression for $\dot{S}_C$, substituting into Eq.~\eqref{eq:second_law}, and using the fact that each individually variable term in the resulting inequality must be nonnegative \cite{coleman_1963,langer_2009a,langer_2009b,langer_2009c}, we arrive at the second-law constraints
\begin{eqnarray}
 \label{eq:W} {\cal W} = V s\, \dot{\gamma}^{\text{pl}} - p \sum_{\alpha} \left( \dfrac{\partial V}{\partial \Lambda_{\alpha}} \right)_{S_C} \dot{\Lambda}_{\alpha} + \dfrac{p\, V\, Y}{\tau} \geq 0; \\
 \label{eq:S_T} (p X - T) \dot{S}_T \geq 0 .
\end{eqnarray}
In arriving at these two constraints, we have arranged terms in such a way that terms pertaining to the degrees of freedom that belong to the same subsystem are grouped together. The dissipation rate ${\cal W}$, as defined in \cite{langer_2009a,langer_2009b,langer_2009c}, is the difference between the rate at which inelastic work is done on the configurational subsystem and the rate at which energy is stored in the internal degrees of freedom. The second constraint implies that $p X - T$ and $\dot{S}_T$ must carry the same sign if they are nonzero, so that
\begin{equation}\label{eq:Q}
 T \dot{S}_T = - V\, {\cal K} (X, T) \left( T - p\, X \right)\equiv \,{\cal Q},
\end{equation}
where ${\cal K} (X, T)$ is a non-negative thermal transport coefficient. It is already clear from this analysis that $p\,X$ plays the role of a temperature. A heat flux ${\cal Q}$ flows between the granular subsystem and the reservoir when the two subsystems are not in thermodynamic equilibrium with each other.

We can now set $T = 0$, and define the dimensionless variables, to be used in the rest of the development, as follows:
\begin{equation}
 \chi \equiv \dfrac{X}{v_Z}; \quad \mu \equiv \dfrac{s}{p} .
\end{equation}
Thus $\chi$ is the dimensionless compactivity, and $\mu$ is the shear-stress-to-pressure ratio. $v_Z$ is the excess volume of STZ's -- loose spots where irreversible particle rearrangements occur, incurring local topological change. The extensive volume $V$ of the granular medium is a sum of the total volume of grains $V_0$ plus the contributions of the configurational degrees of freedom, such as STZ's and misalignments. In the simplest approximation, it varies linearly with $\chi$:
\begin{equation}
 \dfrac{V}{V_0} = 1 + \epsilon_1 (\chi - \chi_r ),
\end{equation}
where $\epsilon_1$ can be interpreted as an effective volumetric expansion coefficient, and $\chi_r$ denotes the compactivity of a reference state.

The equation of motion for $\chi$ is derived from the expression for $\dot{S}_C$. To this end, eliminating $\dot{S}_T$ in Eq.~\eqref{eq:first_law} using Eq.~\eqref{eq:Q} for an expression for $\dot{S}_C$, dividing by $p V_0$, and replacing $\dot{S}_C$ by $\dot{\chi}$ through the chain rule of differential calculus, the equation of motion for $\chi$ becomes
\begin{equation}\label{eq:xdot0}
 \epsilon_1 \dot{\chi} = \mu \dot{\gamma}^{\text{pl}} + \dfrac{Y}{\tau} - {\cal K} (\chi) \chi .
\end{equation}

The next step is to determine the transport coefficient ${\cal K} (\chi)$ in Eq.~\eqref{eq:xdot0}. Because ${\cal K} (\chi)$ couples the configurational and kinetic-vibrational subsystems, we assume that it consists of additive contributions from shearing, interparticle friction, and vibrations. Our strategy is to first determine ${\cal K} (\chi)$ for frictionless grains, then add in the effect of friction afterwards. We know that in the absence of vibrations and interparticle friction, the steady-state compactivity is uniquely determined by the dimensionless shear rate $q \equiv \tau \dot{\gamma}^{\text{pl}}$; that is, $\chi^{\text{ss}} = \hat{\chi}(q)$. As such, direct substitution into Eq.~\eqref{eq:xdot0} in this case yields ${\cal K} = W / (\tau \hat{\chi}(q) )$. Here the dimensionless work rate $W = \tau \mu \dot{\gamma}^{\text{pl}}$ is proportional to the mechanical noise strength $\Gamma$, to be discussed in greater detail in Sec.~\ref{sec:4} below.

On the other hand, if the granular medium is vibrated but not sheared, the steady-state compactivity is determined by the vibration intensity or noise strength $\rho$~\cite{nowak_1998,knight_1995}: $\chi^{\text{ss}} = \tilde{\chi} (\rho)$. As such ${\cal K} = Y / (\tau \tilde{\chi} (\rho) )$. The quantity $\rho$ should be a function of the vibrational frequency and amplitude; past studies~\cite{daniels_2005,daniels_2006,edwards_1998} suggest that $\rho$ is proportional to the vibrational amplitude, and the square of the frequency. However, the detailed dependence is unimportant for the purposes of this paper which, as in the van der Elst \textit{et al.}~experiments~\cite{vanderelst_2012}, focuses on one single amplitude and frequency, unless otherwise specified.

If the granular material is sheared and vibrated at the same time, it is plausible that the shear rate sets a time scale below which vibrations cannot compete with shearing in allowing the system to explore different configurational states by rearranging constituent grains. Referring to Fig.~\ref{fig:xvplot}, and foreshadowing the discussion in the next subsection, the effect of vibrations to increase the packing fraction relative to an unvibrated granular medium diminishes above a certain shear rate. Because the intensity of vibrations is proportional to the square of the frequency~\cite{nowak_1998,knight_1995}, we propose that the diminution of the effect of vibrations is determined by the ratio $q^2 / \rho$. Thus, for smooth frictionless grains, the transport coefficient is
\begin{equation}\label{eq:calK0}
 {\cal K} (\chi) = \dfrac{1}{\tau} \left[ \dfrac{W}{\hat{\chi}(q)} + r \dfrac{Y}{\tilde{\chi}(\rho)} \right] ,
\end{equation}
where $r \equiv \exp (-q^2 / \rho)$ is the relative weight of the contributions from vibrations versus shear. Note that this is for frictionless grains. Of course, $r = 0$ when the granular medium is not vibrated, i.e.~$\rho = 0$. In the cases when the granular medium is subjected to either shear or vibrations but not both, Eq.~\eqref{eq:calK0} reduces to the results discussed in the preceding paragraphs. In analogy to the direct proportionality between the mechanical noise $\Gamma$ and the plastic work of deformation $W$ alluded to above, we propose that the noise strength $\rho$ associated with vibrations is proportional to the work of external vibrations itself: $Y = A_0 \rho$ for some constant $A_0$.

For the case of angular, frictional grains, we propose to modify the coupling coefficient ${\cal K}$ as follows:
\begin{equation}
 {\cal K} (\chi) = \dfrac{1}{\tau} \left[ \dfrac{W + F}{\hat{\chi}(q)} + r \dfrac{Y}{\tilde{\chi}(\rho)} \right] ,
\end{equation}
where $F$ denotes the dissipative effect of friction. As such, the equation of motion for $\chi$, Eq.~\eqref{eq:xdot}, becomes
\begin{equation}\label{eq:xdot2}
 \epsilon_1 \dot{\chi} = \dfrac{W + Y}{\tau} - \dfrac{1}{\tau} \left[ \dfrac{W + F}{\hat{\chi}(q)} + r \dfrac{Y}{\tilde{\chi}(\rho)} \right] \chi .
\end{equation}
The way $F$ appears in the equation ensures that at large shear rates, the dilatational effect of shearing trumps the compaction brought about by external acoustic vibrations and frictional interaction, as suggested by experiments such as~\cite{vanderelst_2012}. This can be seen directly by setting $\dot{\chi} = 0$ in Eq.~\eqref{eq:xdot2} and solving for the steady-state compactivity $\chi^{\text{ss}}$. In analogy to $W$ and $Y$, we propose that $F$ is proportional to what we term the ``frictional noise strength'' $\xi$, to be discussed in Sec.~\ref{sec:5}.

\section{Shear-rate dilation and vibration-driven compaction}\label{sec:3}

At this point let us discuss the quantity $\hat{\chi}(q)$, which denotes the disorder temperature of a driven granular material in the absence of vibrations and interparticle friction. One example of a frictionless granular material is an idealized hard-sphere system. We pointed out in~\cite{lieou_2014b} that this quantity approaches some constant $\hat{\chi}_0$ in the limit of small shear rate $q$, and becomes a rapidly increasing function of $q$ once the shear rate becomes comparable to the rate of intrinsic structural relaxtion, thereby implying shear-rate dilation. A convenient way to interpolate between these two limiting regimes is inspired by the Vogel-Fulcher-Tamann (VFT) form in glass theory~\cite{lieou_2012,haxton_2007,manning_2007b} for the inverse function $q (\hat{\chi})$:
\begin{equation}\label{eq:vft}
 \dfrac{1}{q} = \dfrac{1}{q_0} \exp \left[ \dfrac{A}{\hat{\chi}} + \alpha_{\text{eff}} (\hat{\chi}) \right] ,
\end{equation}
where
\begin{equation}\label{eq:alphaeff}
 \alpha_{\text{eff}} (\hat{\chi}) = \left( \dfrac{\hat{\chi}_1}{\hat{\chi} - \hat{\chi}_0} \right) \exp \left( - 3 \dfrac{\hat{\chi} - \hat{\chi}_0}{\hat{\chi}_A - \hat{\chi}_0} \right) .
\end{equation}
In the rest of the paper, we have chosen $\hat{\chi}_0 = 0.3$, $\hat{\chi}_1 = 0.02$, $\hat{\chi}_A = 0.33$, $A = 2$, and $q_0 = 2$; these values are identical to those used in~\cite{lieou_2014b} to analyze experimental data for auto-acoustic compaction~\cite{vanderelst_2012} at intermediate shear rates. (The appearance of the constant 3 in Eq.~\eqref{eq:alphaeff} is purely a matter of convention.)

The choice $A = 2$ in Eq.~\eqref{eq:vft} stipulates a rate-strengthening response of the granular material to applied shear, at least in the large $q$ regime. An earlier analysis by Daub and one of us~\cite{daub_2009} involved a model for which $A < 1$, which corresponds to rate-weakening behavior; stick-slip instabilities did occur in that case. However, experience shows that hard-sphere systems, or ``barebone'' granular materials, should be rate strengthening~\cite{lieou_2012}. A key result of this paper is that the condition $A < 1$ can be lifted in the presence of other microscopic mechanisms, such as interparticle friction, that alter the hard-sphere rheology, allowing for the possibility of stick-slip. In other words, we isolate the mechanism for rate-weakening behavior and stick-slip instabilities in a naturally-occurring granular medium.

Meanwhile, the quantity $\tilde{\chi} (\rho)$, which denotes the steady-state compactivity of a vibrated granular medium, should be a decreasing function of increasing vibration intensity $\rho$. (As we pointed out above, $\rho$ is a function of the vibration frequency and amplitude.) This is because an increase in vibration intensity causes the grains to explore more possible configurations and reduce the packing fraction~\cite{nowak_1998,knight_1995}, dissipating energy in the process. Since we consider only one vibration intensity, we do not propose a functional form for $\tilde{\chi} (\rho)$, but stipulate that it takes on one of two values depending on whether the fixed-intensity vibration is turned on or off. The combined effect of shearing, vibrations, and friction on the steady-state compactivity $\chi^{\text{ss}}$ is illustrated in Fig.~\ref{fig:xvplot} above for vibrated and unvibrated granular media composed of frictional and frictionless particles.

\section{Microscopic model of STZ's}\label{sec:4}

We now turn to the Shear-Transformation-Zone (STZ) theory of dense granular flow~\cite{lieou_2012,falk_1998,langer_2011,lieou_2014a}, which directly connects macroscopic dynamics to the grain-scale physics of deformation. The latter refers to the irreversible, nonaffine granular rearrangements, along with noise induced by friction and external vibration. The STZ theory has been invoked in our successful effort to reconstruct and explain constitutive friction laws~\cite{daub_2010,lieou_2014a,elbanna_2014}.

Recall our physical picture that in dense granular flow, irreversible particle rearrangements occur at rare, non-interacting soft spots with excess free volume known as STZ's. The applied shear stress defines a direction relative to which STZ's can be classified according to orientation, with total numbers $N_+$ and $N_-$ respectively. Upon application of shear stress in the ``plus'' direction, STZ's of the minus type easily deform to become plus-type STZ's. However, plus-type STZ's rarely flip and acquire the minus orientation; rather, they are annihilated readily by noise. This is described by a master equation of the form
\begin{equation}\label{eq:master}
 \tau \dot{N}_{\pm} = {\cal R} (\pm \mu, \chi) N_{\mp} - {\cal R} (\mp \mu, \chi) N_{\pm} + \tilde{\Gamma} \left( \dfrac{1}{2} N^{\text{eq}} - N_{\pm} \right) .
\end{equation}
Here, we assume that the attempt frequency $\tilde{\Gamma} = \Gamma + \rho$, being the sum of mechanical and vibrational noise strengths, does not include contributions from $\xi$, because interparticle friction serves to dissipate energy but does not open up or close voids. The inertial time scale $\tau = a \sqrt{ \rho_G / p}$~\cite{jop_2006}, with $a$, $\rho_G$, and $p$ being the characteristic grain size, grain material density, and confining pressure, is the typical duration for a pressure-driven granular rearrangement.

If $N$ is the total number of grains, the plastic strain rate is
\begin{equation}\label{eq:strainrate}
 \dot{\gamma}^{\text{pl}} = \dfrac{2\,\epsilon_0}{\tau N} \left[ {\cal R}(\mu, \chi) N_- - {\cal R}(-\mu, \chi) N_+ \right].
\end{equation}
Introducing the intensive variables
\begin{equation}
 \Lambda = \dfrac{N_+ + N_-}{N}; \quad m = \dfrac{N_+ - N_-}{N_+ + N_-},
\end{equation}
which denote the density and orientational bias of STZ's, and analogously the rate factors
\begin{eqnarray}
 \label{eq:calC}{\cal C}(\mu, \chi) &=& \dfrac{1}{2} \left( {\cal R}(\mu, \chi) + {\cal R}(-\mu, \chi) \right) ; \\
 \label{eq:calT}{\cal T}(\mu, \chi) &=& \dfrac{{\cal R}(\mu, \chi) - {\cal R}(-\mu, \chi)}{{\cal R}(\mu, \chi) + {\cal R}(-\mu, \chi)} ,
\end{eqnarray}
Eqs.~\eqref{eq:master} and \eqref{eq:strainrate} can be rewritten in terms of intensive quantities alone:
\begin{eqnarray}
 \label{eq:Lambda} \tau\, \dot{\Lambda} &=& \tilde{\Gamma} ( \Lambda^{\text{eq}} - \Lambda ) ; \\
 \label{eq:m} \tau\, \dot{m} &=& 2\, {\cal C}(\mu, \chi) ( {\cal T}(\mu, \chi) - m ) - \tilde{\Gamma} m - \tau \dfrac{\dot{\Lambda}}{\Lambda} m ; ~~~~~ \\
 \label{eq:D_pl} \tau \,\dot{\gamma}^{\text{pl}} &=&2\, \epsilon_0\,\Lambda\, {\cal C}(\mu, \chi) ({\cal T}(\mu, \chi) - m),
\end{eqnarray}

Note that $\Lambda$ and $m$ are not the only internal state variables; others include the orientational bias of angular grains~\cite{lieou_2014b}. However, at least for grains with aspect ratio close to unity, those variables are not coupled to Eq.~\eqref{eq:D_pl}, and therefore play no role in plastic shear deformation. Thus, internal variables other than $\Lambda$ and $m$ are expected to be immaterial in the present analysis, in which we are not explicitly concerned with the volumetric response of the system.

We now substitute Eqs.~\eqref{eq:Lambda}, \eqref{eq:m}, and \eqref{eq:D_pl} into Eq.~\eqref{eq:W} for the dissipation rate ${\cal W}$ which, according to the second law of thermodynamics, must be non-negative. We then arrive at several independent constraints on the dynamics of the system, according to the procedure outlined in~\cite{lieou_2012,lieou_2014b,coleman_1963,langer_2009a,langer_2009b,langer_2009c}. One of these constraints says that the STZ density at equilibrium equals $\Lambda^{\text{eq}} = 2 e^{-1 / \chi}$. Another tells us that ${\cal T}(\mu, \chi) = \tanh ( \epsilon_0 \mu / \epsilon_Z \chi )$, where $\epsilon_Z$ is the excess volume $v_Z$ of each STZ normalized by the grain size $a^3$.

\subsection{Mechanical noise and steady-state dynamics}

Since STZ's are rare and non-interacting, $\Lambda \ll 1$. Observe that the equations of motion for $\Lambda$ and $m$, Eqs.~\eqref{eq:Lambda} and \eqref{eq:m}, do not contain the small factor $\Lambda$, but Eq.~\eqref{eq:D_pl} does. This provides the basis for a further simplification -- the quasisationary approximation -- under which we can replace $\Lambda$ and $m$ in the equations by their quasistationary values $\Lambda^{\text{eq}}$ and $m^{\text{eq}}$. We have argued that $\Lambda^{\text{eq}} = 2 e^{-1 / \chi}$. To calculate $m^{\text{eq}}$, invoke the Pechenik hypothesis, which states that the mechanical noise strength $\Gamma$ is proportional to the plastic work of deformation dissipated per STZ~\cite{pechenik_2003}:
\begin{equation}\label{eq:Gamma}
 \Gamma = \dfrac{\tau \mu \dot{\gamma}^{\text{pl}}}{\epsilon_0 \mu_0 \Lambda^{\text{eq}}} = \dfrac{2 \mu}{\mu_0} R_0 \Bigl( {\cal T}(\mu, \chi) - m \Bigr).
\end{equation}
In the last equality we replaced ${\cal C}(\mu, \chi)$ by some constant $R_0$, for the STZ flipping rate should not depend strongly on the shear stress as long as the shear-stress-to-pressure ratio $\mu$ is less than unity. Then the stationary version of Eq.~\eqref{eq:m} reads
\begin{equation}
 2\, R_0 \Bigl( {\cal T}(\mu, \chi) - m \Bigr) \left( 1 - \dfrac{m \mu}{\mu_0} \right) - m\, \rho  = 0 ,
\end{equation}
and the stationary value of $m$ is given by
\begin{eqnarray}\label{eq:m_eq}
 \nonumber m^{\text{eq}} &=& \dfrac{\mu_0}{2\mu} \left[ 1 + \dfrac{\mu}{\mu_0} {\cal T}(\mu, \chi) + \dfrac{\rho}{2 R_0} \right] \\ &&- \dfrac{\mu_0}{2\mu} \sqrt{\left[ 1 + \dfrac{\mu}{\mu_0} {\cal T}(\mu, \chi) + \dfrac{\rho}{2 R_0} \right]^2 - \dfrac{4 \mu}{\mu_0} {\cal T}(\mu, \chi)} . ~~~~~
\end{eqnarray}
In particular, when $\rho = 0$, we have
\begin{equation}
 m^{\text{eq}} = 
\begin{cases}
{\cal T}(\mu, \chi), & \text{if $(\mu / \mu_0)\, {\cal T}(\mu, \chi) < 1$} ; \\
\mu_0 /\mu, & \text{if $(\mu / \mu_0)\, {\cal T}(\mu, \chi) \geq 1$}.
\end{cases}
\end{equation}
Thus, in effect, $\mu_0$ sets the minimum flow stress of the system in the absence of external vibrations. However, when $\rho \neq 0$, the system is un-jammed, and flows at arbitrarily small shear stress $s$. 

\subsection{Yield stress parameter and apparent weakening}

\begin{figure}[ht]
\centering 
\includegraphics[width=.45\textwidth]{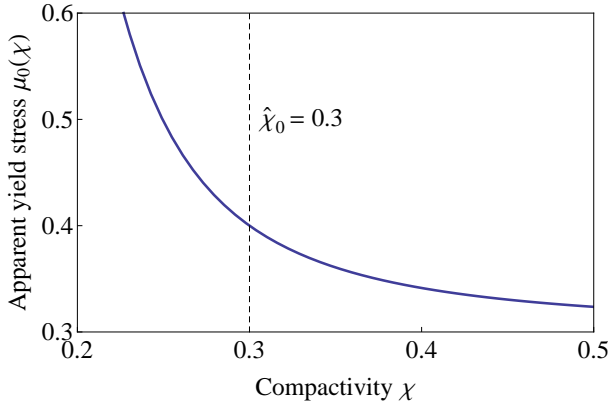}
\caption{\label{fig:s0plot}(Color online) The parameter $\mu_0$, which provides a measure of the apparent yield stress, as a function of the compactivity $\chi$. This quantity must be a rapidly decreasing function of $\chi$ whenever $\chi < \hat{\chi}_0$, and cease to be so above $\hat{\chi}_0$, as indicated by the vertical dashed line. ($\hat{\chi}_0$ is the steady-state compactivity in the slow shear limit.) This condition conforms with the intuition that the stress needed to unjam a granular medium increases with the packing fraction. This, combined with the rapid increase of the steady-state compactivity $\chi^{\text{ss}}$ below $\hat{\chi}_0$ for frictional grains shown in Fig.~\ref{fig:xvplot}, provides the apparent weakening that permits stick-slip instabilities at high enough pressure. Here we have used the interpolation Eq.~\eqref{eq:mu_0} with $\mu_1 = 0.3$ and $\mu_2 = 3.7 \times 10^{-3}$.} 
\end{figure}

The yield stress parameter $\mu_0$, which originates as a proportionality parameter between the mechanical noise strength $\Gamma$ and the plastic work of deformation, plays a crucial role in the present analysis. In~\cite{lieou_2012} we assumed that $\mu_0$ is a rapidly decreasing function of the kinetic-vibrational temperature below the glass transition temperature. Yet it could equally well be a decreasing function of the compactivity $\chi$ if we extend the analogy between the thermal and effective temperatures. There are two ways to understand this. First, intuition suggests that the amplitude of the fluctuations generated by shearing decreases as the granular medium becomes more tightly packed. Thus $\mu_0$ should decrease with increasing $\chi$. Second, one expects that a larger shear stress is needed to create voids in a more densely-packed granular medium in order for it to flow. In this connection, we pointed out above that $\mu_0$ controls the minimum flow stress in the absence of vibrations. Indeed, a $\mu_0$ that decreases rapidly with $\chi$ when $\chi$ is below the steady-state compactivity $\hat{\chi}_0$ in the slow-shear limit, such as shown in Fig.~\ref{fig:s0plot}, is an essential ingredient for the emergence of stick-slip instabilities here.

On the other hand, $\mu_0$ must cease to be rapidly decreasing when $\chi > \hat{\chi}_0$, and be bounded from below by a nonzero value, so that the energy dissipated by shear-induced fluctuations (to open up voids and create STZ's, for example) cannot exceed the energy fed into the granular medium itself. Thus we use
\begin{equation}\label{eq:mu_0}
 \mu_0 = \mu_1 + \mu_2 e^{1 / \chi}
\end{equation}
to interpolate between the large- and small-$\chi$ regimes. We argue below that a $\mu_0$ with these qualitative features is one of two ingredients that provide apparent rate-weakening at intermediate shear rates. The other ingredient is the nonmonotonic variation of the steady-state compactivity $\chi^{\text{ss}}$ with shear rate, shown in Fig.~\ref{fig:xvplot}. Referring to Eq.~\eqref{eq:xdot2} and the subsequent discussion, this is provided by appropriate noise strengths $\rho$ and $\xi$ associated with vibrations and interparticle friction. The apparent weakening shows up in the phase plot of shear stress $s$ versus plastic strain rate $\dot{\gamma}^{\text{pl}}$ in Fig.~\ref{fig:phaseplot}.

\section{Emergence of stick-slip instabilities}\label{sec:5}

\subsection{Equations of motion}

The complete constitutive description consists of two dynamical equations, one for the shear-stress-to-pressure ratio $\mu$ and one for the compactivity $\chi$. The equation for $\mu$ is a statement of linearity between increments in stress and elastic strain. Thus,
\begin{equation}\label{eq:sdot}
 \dot{\mu} = (G / p) (\dot{\gamma} - \dot{\gamma}^{\text{pl}} ),
\end{equation}
where $G$ is the aggregate shear modulus, and $\dot{\gamma}$ is the imposed shear rate. The plastic strain rate is given directly by Eq.~\eqref{eq:D_pl}:
\begin{equation}\label{eq:v_pl}
 \dot{\gamma}^{\text{pl}} = \dfrac{4 \epsilon_0}{\tau} e^{-1 / \chi} R_0 \left[ {\cal T}(\mu, \chi) - m (\mu, \chi) \right] .
\end{equation}
Meanwhile, the equation for the disorder temperature $\chi$ describes how the fluctuations $\Gamma$ induced by shearing, and $\rho$ induced by external vibrations, drive the system towards different configurational steady states, described respectively by $\hat{\chi}(q)$ and $\tilde{\chi}(\rho)$. Additionally it describes how friction-induced noise $\xi$ when the material is sheared results in further dissipation and compaction:
\begin{eqnarray}\label{eq:xdot}
 \nonumber \dot{\chi} &=& \dfrac{2 \epsilon_0 \mu_0 e^{-1 / \chi}}{\tau \epsilon_1} \left[ \Gamma \left( 1 - \dfrac{\chi}{\hat{\chi}(q)} \right) - \xi \dfrac{\chi}{\hat{\chi}(q)} \right] \\ & &+ \dfrac{A_0 \rho}{\tau \epsilon_1} \left[ 1 - \exp \left(- \dfrac{q^2}{\rho} \right) \dfrac{\chi}{\tilde{\chi} (\rho)} \right] ~~~~~.
\end{eqnarray}
To derive Eq.~\eqref{eq:xdot} from Eq.~\eqref{eq:xdot2} directly, it suffices to identify in Eq.~\eqref{eq:xdot2} $W = \Gamma \epsilon_0 \mu_0 \Lambda$, $Y = A_0 \rho$, and $F = \xi \epsilon_0 \mu_0 \Lambda$. The first two of these equalities have been discussed in connection with Eq.~\eqref{eq:calK0}; they imply that the work rates of shearing and vibrations are proportional to the respective noise strengths. The last equality connects the frictional dissipation $F$ with the noise strength $\xi$ in the same manner. It simply says that frictional dissipation increases with the number of STZ's present in the granular medium. We discuss in the following subsection the noise strength $\xi$ in greater detail.

\subsection{Noise strength $\xi$ associated with interparticle friction}

The steady-state solution to Eq.~\eqref{eq:xdot} shows that if $\xi$ varies as $(\dot{\gamma}^{\text{pl}})^2$ at small shear rates, and saturates to some constant $\xi_0$ at large $\dot{\gamma}^{\text{pl}}$, then it is possible for $\chi$ to decrease below the zero-shear-rate limit $\hat{\chi}_0$ for a range of intermediate shear rates. This is illustrated in Fig.~\ref{fig:xvplot}. We argued in~\cite{lieou_2014b} that this nonmonotonic variation of $\chi$ explains the occurrence of auto-acoustic compaction in that regime. Thus we employ the interpolation 
\begin{equation}\label{eq:xi}
 \xi = \xi_0 \tanh [(\tau_f \dot{\gamma}^{\text{pl}})^2]
\end{equation}
between the two limiting regimes with $\tau_f$ being a time scale at which the compactional effect of interparticle friction becomes prominent.

At somewhat larger shear rates, $\chi$ increases as a function of the shear rate but remains smaller than $\hat{\chi}_0$. ($\chi$ only exceeds $\hat{\chi}_0$ and diverges at very fast shear rates.) Here, $\mu_0$ is a rapidly decreasing function of increasing $\chi$. As a result, $\mu_0$ decreases markedly as $\dot{\gamma}^{\text{pl}}$ increases. This apparent rate-weakening behavior, summarized in Figs.~\ref{fig:xvplot} and~\ref{fig:s0plot}, permits stick-slip instabilities in unvibrated frictional granular materials for this range of intermediate shear rates. One might ask whether stick-slip could also occur for unvibrated frictionless particles, which correspond to the dashed red curve in Fig.~\ref{fig:xvplot}, for which there is also a rapid increase of $\chi$ as a function of $\dot{\gamma}^{\text{pl}}$ even when $\chi < \hat{\chi}_0$. However, vibrations fluidize the granular medium and cause weakening. That is, a vibrated granular material can experience irreversible deformation at arbitrarily small stresses, and the behavior of $\mu_0$ becomes immaterial in that regime. There is no physical mechanism that brings the steady-state $\chi$ below $\hat{\chi}_0$ and then up again. Therefore stick-slip instabilities do not occur in that case.

\subsection{Results}

We now present the key predictions and implications of the STZ granular flow model. Equations~\eqref{eq:sdot} and \eqref{eq:xdot} are numerically integrated using an adaptive, implicit time-stepping scheme, for a range of values of the pressure $p$ and imposed shear rate $\dot{\gamma}$, and initial values $\mu(t = 0) = 10^{-4}$ and $\chi(t = 0) = 0.18$. We focus on angular quartz sand with a typical particle diameter of $a = 350$ $\mu$m, mass density $\rho_G = 1600$ kg m$^{-3}$, and aggregate shear modulus $G = 109$ MPa. Because of the complexities in the functional dependencies of the various parameters, we have chosen not to use dimensionless parameters in our results. Instead, we express the shear rate in units of s$^{-1}$, and pressure in units of MPa, to make the pressure-dependent behavior more transparent. We take $R_0 = 1$, $A_0 = 0.01$, $\epsilon_0 = 1.5$, $\epsilon_Z = 0.5$, and $\epsilon_1 = 0.3$ in Eqs.~\eqref{eq:v_pl} and \eqref{eq:xdot}, and $\xi_0 = 4 \times 10^{-3}$, $\tau_f = 0.013$ s, and $\hat{\chi}_0 = 0.3$. For comparison we also compute the response of frictionless spherical beads to shear, with $\xi_0 = 0$ and otherwise identical material parameters. Referring to Sec.~\ref{sec:3} above, our choice of $\hat{\chi}(q)$ automatically stipulates rate-strengthening rheology in the limit of large $\dot{\gamma}^{\text{pl}}$. Unless otherwise specified, we take $\tilde{\chi} (\rho = 0) = 0.2$ in the absence of external acoustic vibrations, and $\tilde{\chi} (\rho) = 0.18$ and $\rho = 5 \times 10^{-4}$ in its presence. Most of these parameters are identical to those used in~\cite{lieou_2014b} to model auto-acoustic compaction at intermediate shear rates. The parameter values are either determined empirically, such as for $\epsilon_1$ and $\epsilon_Z$, or are chosen to quantitatively describe the experimental findings of \cite{vanderelst_2012}. In addition, we choose $\mu_1 = 0.3$ and $\mu_2 = 3.7 \times 10^{-3}$, so that $\mu_0 (\chi)$ decreases rapidly as a function of $\chi$ when $\chi < \hat{\chi}_0 = 0.3$, and levels off above $\hat{\chi}_0$. 

\begin{figure}[ht]
\centering 
\includegraphics[width=.45\textwidth]{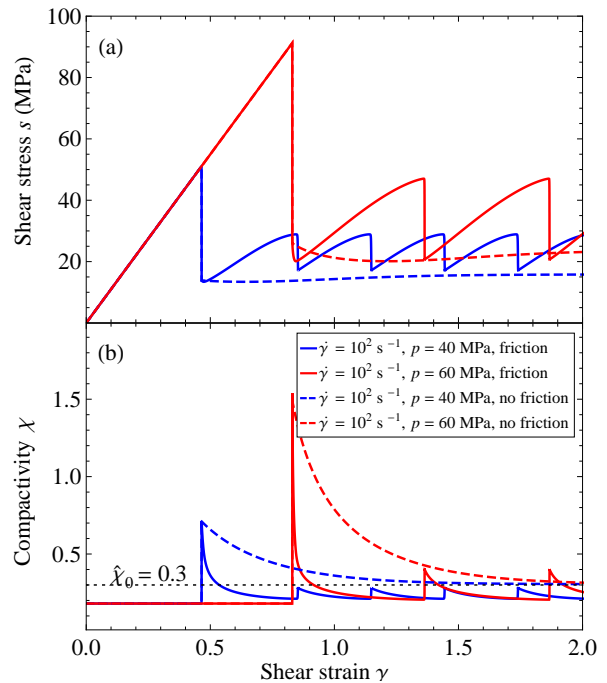}
\caption{\label{fig:nofrplot}(Color online) Variation of (a) shear stress $s$ and (b) compactivity $\chi$ with shear strain, in the presence (solid curves) and absence (dashed curves) of interparticle friction, and in the absence of external vibration. Observe that stick-slip occurs only in the presence of interparticle friction, which provides a means for the compactivity to fall below $\hat{\chi}_0 = 0.3$, shown with the dotted line in (b), in the ideal hard-sphere model, and facilitates the emergence of instabilities. The slip phase of stick-slip coincides with a sharp increase in the compactivity, or volume dilatation.} 
\end{figure}

Figure~\ref{fig:nofrplot} makes transparent the essential role of interparticle friction in the emergence of stick-slip. No stick-slip occurs in the absence of friction, that is, when $\xi = 0$ in Eq.~\eqref{eq:xdot}. (For clarity we plot the shear stress $s$ instead of the shear stress to pressure ratio $\mu$.) During the initial loading, the red and blue curves for stress versus strain in Fig.~\ref{fig:nofrplot}(a) describe almost purely elastic deformation; thus $\chi$ remains constant at its initial value. This is independent of the presence of interparticle friction. Then, when the stress exceeds the yield stress for that value of $\chi$, the plastic strain rate $\dot{\gamma}^{\text{pl}}$ jumps up, as does $\chi$ according to Eq.~\eqref{eq:xdot}. A dramatic stress drop follows immediately, as a consequence of Eq.~\eqref{eq:sdot}, causing the arrest of slip. In the absence of interparticle friction, the compactivity $\chi$ remains large, so that $\mu_0 (\chi)$ stays small. As such, the material remains soft, and no further slipping occurs.

However, this may not be the case if there is friction. As we argued above and in~\cite{lieou_2014b}, the noise $\xi$ associated with interparticle friction dissipates energy and cools down the system in the volumetric sense; it causes the compactivity $\chi$ to fall below $\hat{\chi}_0 = 0.3$, seen also in the solid curves in Figs.~\ref{fig:xvplot} and~\ref{fig:nofrplot}(b). This is a regime where, according to Fig.~\ref{fig:s0plot}, $\mu_0 (\chi)$ is a rapidly decreasing function of $\chi$, providing the necessary weakening effect for stick-slip instabilities to occur. Note the important role played by the $\chi$-dependence of the apparent yield stress $\mu_0 (\chi)$. Interparticle friction increases the shear stress at all times, conforming with the notion that friction increases the difficulty for slip to occur. Note also that the slip phase corresponds to a dramatic increase in the compactivity $\chi$, which increases monotonically with the layer thickness or volume~\cite{lieou_2014b}. A direct connection to laboratory measurements of volume or layer thickness can be made once the quantitative relation between $\chi$ and the volume is established. Thus the disorder temperature $\chi$ provides a natural link between stick-slip dynamics, volumetric effects, and energy dissipation.

\begin{figure}[t!]
\centering 
\includegraphics[width=.45\textwidth]{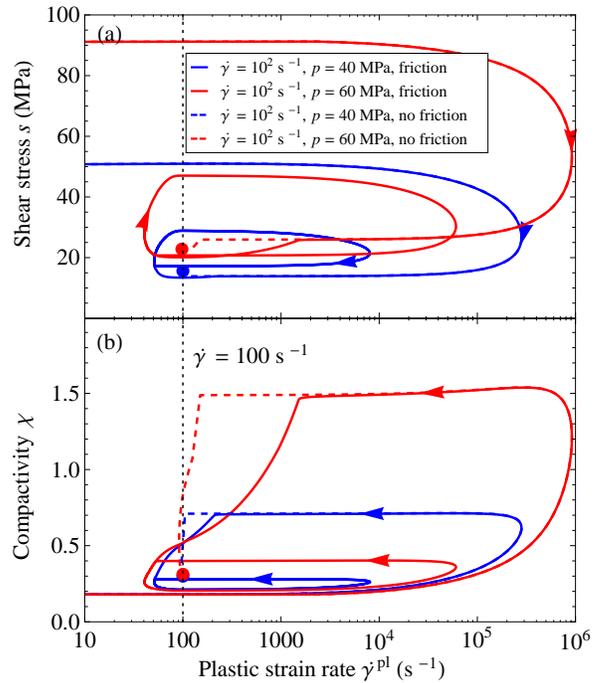}
\caption{\label{fig:phaseplot}(Color online) Variation of (a) shear stress $s$ and (b) compactivity $\chi$ with the plastic strain rate $\dot{\gamma}^{\text{pl}}$ in the presence (solid curves) and absence (dashed curves) of interparticle friction, and in the absence of external vibration. The pressure values are $p = 40$ MPa (blue) and $60$ MPa (red), and the imposed strain rate is $\dot{\gamma} = 10^2$ s$^{-1}$, indicated by the vertical dotted line. Stick-slip occurs for frictional grains, and each of the corresponding phase curves converges to a limit cycle. For frictionless grains, each of the phase curves converges to a fixed point, indicated by each of the filled circles, on the phase diagram. (The fixed points of $\chi$ versus $\dot{\gamma}^{\text{pl}}$ for $p = 40$ and $60$ MPa overlap with one another.) In the case of frictional grains, the applied load ``sticks'' left of the vertical dotted line, and creeps at a slower but nonzero rate than the imposed strain rate, while the compactivity decreases. Once the shear stress exceeds that needed to overcome frictional resistance and initiate slip, both the compactivity and the plastic strain rate increase dramatically, and the stress drops. Then the applied load sticks, and the shear stress builds up again during the new stick phase. Thus the stick-slip cycle repeats itself. This representation of stick-slip dynamics can be directly compared to that in Fig.~\ref{fig:nofrplot}.}
\end{figure}

Another way to visualize the dynamics is to construct a phase plot for the evolution of shear stress and compactivity, as functions of the plastic strain rate $\dot{\gamma}^{\text{pl}}$, in each stick-slip cycle. This is shown in Fig.~\ref{fig:phaseplot}, where the imposed strain rate $\dot{\gamma}$ is indicated by the vertical dotted line. In each case the material starts from nearly zero shear stress and is loaded elastically; thus $\dot{\gamma}^{\text{pl}} = 0$ and is not shown on the phase plot. Then, when the shear stress exceeds the yield stress, plastic deformation sets in; $\dot{\gamma}^{\text{pl}}$ increases dramatically, depicted by the horizontal portions of the phase plots shooting towards the right. The release of plastic strain results in a dramatic stress drop, after which the the phase curve either converges to a fixed point in the case of frictionless grains, or a limit cycle in the case of stick-slip in frictional grains, for the values of pressure $p$ and imposed strain rate $\dot{\gamma}$ shown here. Creep occurs during the stick phase; that is, plastic deformation occurs at a much slower, but nonzero, rate than the imposed strain rate. The shear stress increases, and the compactivity decreases, so that the granular layer contracts, until the applied load overcomes the frictional resistance. This is followed by a dramatic increase of the plastic strain rate, or slip rate, which corresponds to the compactivity surge and stress drop. Because of the elasticity of the granular medium itself, the applied load ``sticks'' again, and the stick-slip cycle repeats itself. The apparent weakening behavior is evident from Fig.~\ref{fig:phaseplot}(a); the stick-slip phase curves always traverse the limit cycles in the clockwise sense. Stress drops occur at large plastic strain rates, while stress increases develop at lower $\dot{\gamma}^{\text{pl}}$.

\begin{figure}[t!]
\centering 
\includegraphics[width=.45\textwidth]{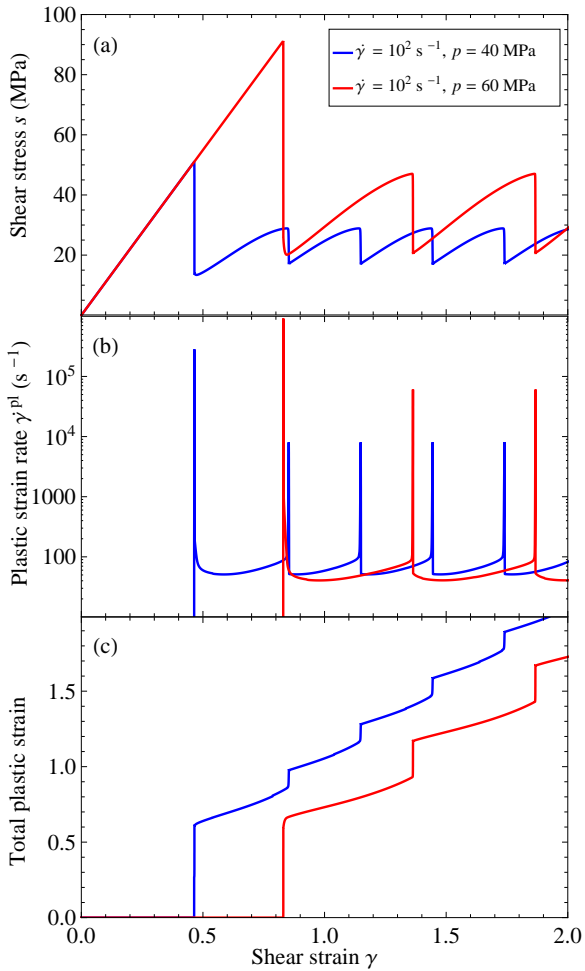}
\caption{\label{fig:pssplot}(Color online) Variation of (a) shear stress $s$, (b) plastic strain rate $\dot{\gamma}^{\text{pl}}$, and (c) accumulated plastic strain with total shear strain, in sheared, unvibrated angular sand, a granular material composed of frictional grains. Observe from (b) that $\dot{\gamma}^{\text{pl}}$ is slowly increasing during the stick phase of each stick-slip cycle; this accounts for the the apparent softening seen towards the end of each stick-slip cycle in (a). Panel (c) shows that while much of the plastic strain is accumulated during the slip phase, there is plastic strain during the stick phase as well.}
\end{figure}

Of interest is the rounding of the stress-strain curves near each slip event; that is, in each stick-slip cycle, the rate of increase of the shear stress $s$ during the stick phase decreases with increasing shear strain. This creep is also seen in experiments~\cite{johnson_2008}. Referring to Eq.~\eqref{eq:sdot}, the plastic strain rate $\dot{\gamma}^{\text{pl}}$ must therefore be nonzero even during the stick phase. This is reminiscent of preseismic slip in the context of earthquakes; that is, small but nonzero plastic deformation prior to the large slip events that correspond to dramatic stress drop. Evidence of this is shown in Fig.~\ref{fig:pssplot}. Panel (a) shows once again the stress-strain curves. In panel (b), we plot the plastic strain rate $\dot{\gamma}^{\text{pl}}$ versus the total shear strain $\gamma$, and show that the plastic strain rate is nonzero and increasing even during the stick phase of each stick-slip cycle. During the slip phase, $\dot{\gamma}^{\text{pl}}$ increases very rapidly, and peaks at a value that is a few orders of magnitude above the imposed strain rate $\dot{\gamma}$. As a result of the stress drop, the plastic strain rate plummets almost instantaneously and the load sticks again. We plot in panel (c) the accumulated plastic strain
\begin{equation}
 \gamma^{\text{pl}} \equiv \int_0^t dt' \dot{\gamma}^{\text{pl}} (t')
\end{equation}
as a function of total shear strain $\gamma$, and show that while much of the plastic strain is accumulated during the slip phase of each stick-slip cycle, there is indeed a slow build-up of plastic strain, or creep, during the stick phase. This corroborates with the result shown in Fig.~\ref{fig:pssplot}(b). While not shown here, the same qualitative behavior is observed in vibrated frictional sand particles when stick-slip occurs.


\begin{figure}[t!]
\centering 
\subfigure{\includegraphics[width=.45\textwidth]{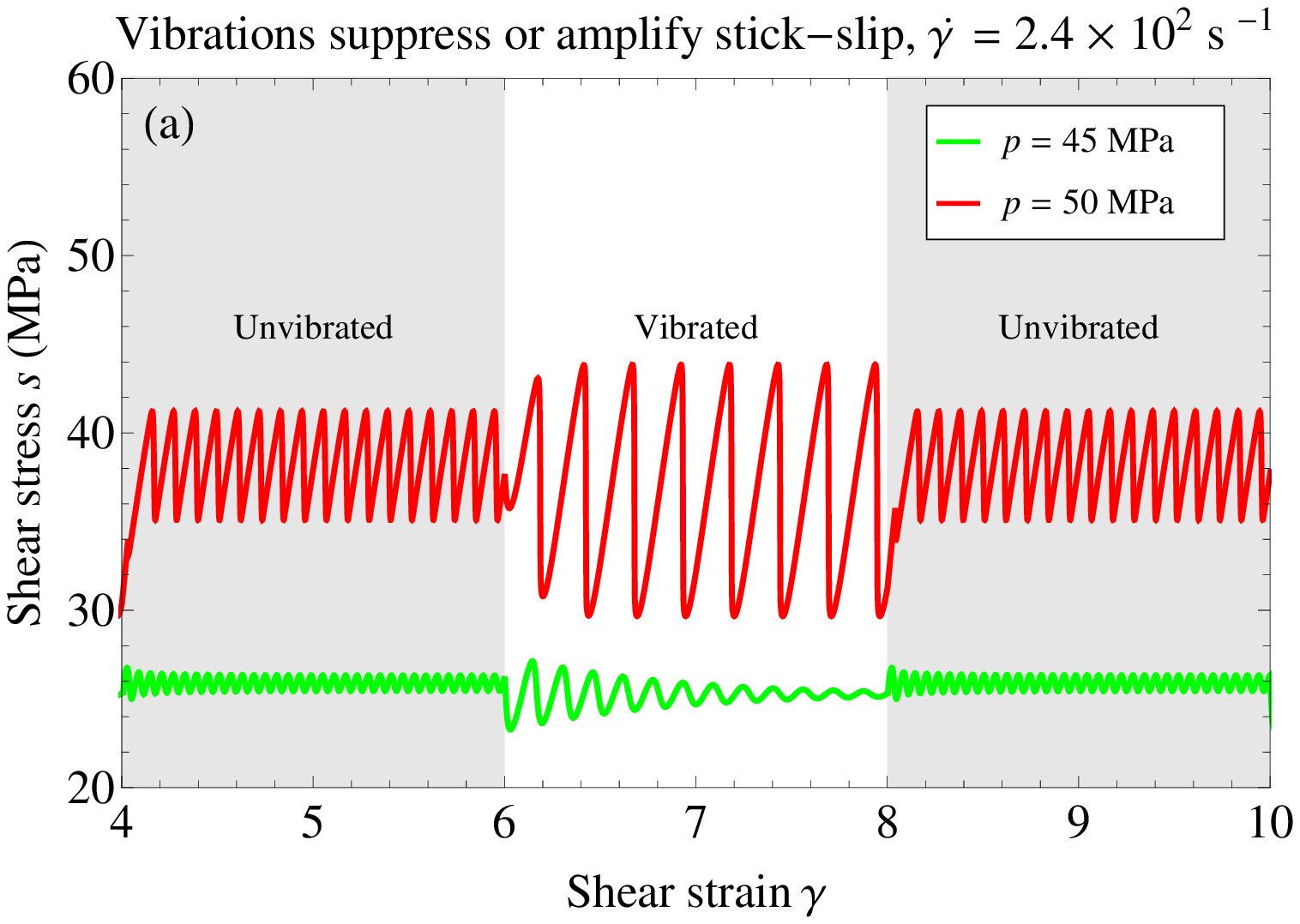}\label{fig:zsplot}}
\subfigure{\includegraphics[width=.45\textwidth]{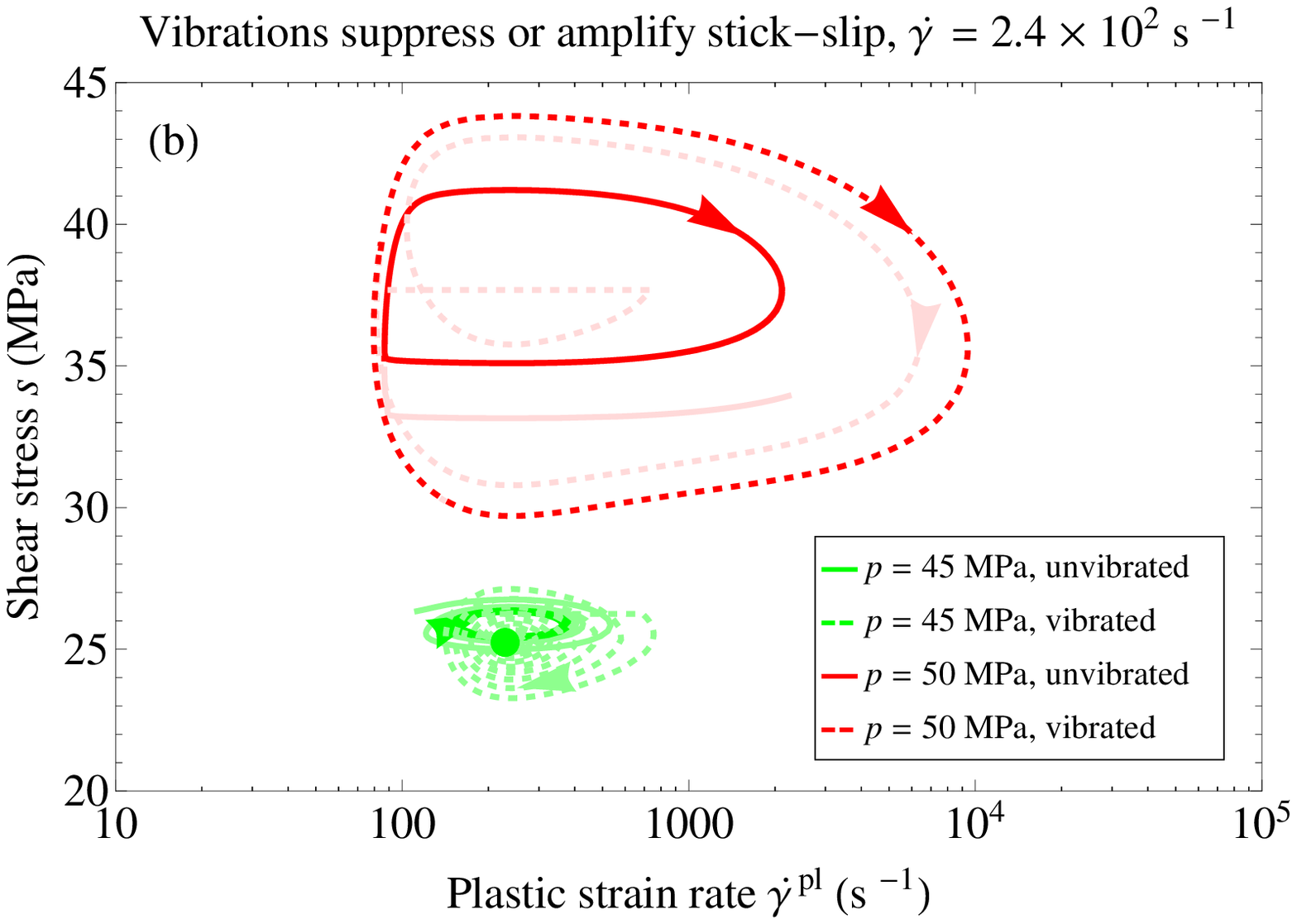}\label{fig:zphaseplot}}
 \caption{\label{fig:sxplot}(Color online) Vibrations suppress or amplify stick-slip in different parameter regimes. (a) Shear stress $s$ a function of shear displacement $\gamma$ for sheared angular sand, for different magnitudes of the normal stress $p$, at applied shear rate $\dot{\gamma} = 2.4 \times 10^2$ s$^{-1}$. External acoustic vibration is turned on (with intensity $\rho = 5 \times 10^{-4}$) for $6 < \gamma < 8$, and off otherwise. At the lower pressure of $p = 45$ MPa (green curve), vibrations amplify stick-slip amplitude before atttenuating it. At the higher pressure of $p = 50$ MPa (red curve), vibrations amplify the stick-slip amplitude and period. The red curve for $p = 50$ MPa has been offset by $10$ MPa for clarity. (b) Variation of shear stress $s$ with plastic shear rate $\dot{\gamma}^{\text{pl}}$ under the conditions in (a) above, for shear strains $4 < \gamma < 8$. The dashed lines denote the behavior under vibrations of fixed intensity, while the lighter shades indicate transients before approaching a limit cycle or fixed point. At the lower pressure of $p = 45$ MPa, except for a short transient the phase curve traces a limit cycle when the sand is not vibrated. The phase curve spirals towards a single fixed point, denoted by the filled green circle, when the material is vibrated. At the higher pressure of $p = 50$ MPa, except for short transients immediately following the switching on or off of fixed-intensity external vibrations, the phase curve traces a smaller limit cycle in the clockwise direction when the material is not subject to vibrations, and a bigger limit cycle when vibrations are switched on.} 
\end{figure}

Figures~\ref{fig:zsplot} shows sample stress-strain response for angular sand, with external vibrations turned on and off alternately. Here, the imposed shear rate is $\dot{\gamma} = 2.4 \times 10^2$ s$^{-1}$, and the pressure values are $p = 45$ and $50$ MPa. Several interesting features are observed. First, at the lower pressure of $p = 45$ MPa, vibrations amplify the stick-slip amplitude before attenuating it, and steady sliding takes over at long enough times. Second, at the higher pressure of $p = 50$ MPa, vibrations amplify the stick-slip amplitude and period. These observations are depicted in another manner in Fig.~\ref{fig:zphaseplot} where we show the variation of shear stress $s$ with plastic strain rate $\dot{\gamma}^{\text{pl}}$ in a phase plot. The sudden onset or ceasing of vibrations results in a short transient in both cases, denoted by lighter shades. Subsequently the phase curve of $s$ versus $\dot{\gamma}^{\text{pl}}$ approaches and traces a limit cycle, or spirals towards a fixed point in the case of vibrated sand at $p = 45$ MPa. The limit cycle provides a straightforward illustration of the stick-slip amplitude. In other words, vibrations can either change the stick-slip amplitude or suppress stick-slip, depending on the other control parameters.

\begin{figure}[ht]
\centering 
\includegraphics[width=.45\textwidth]{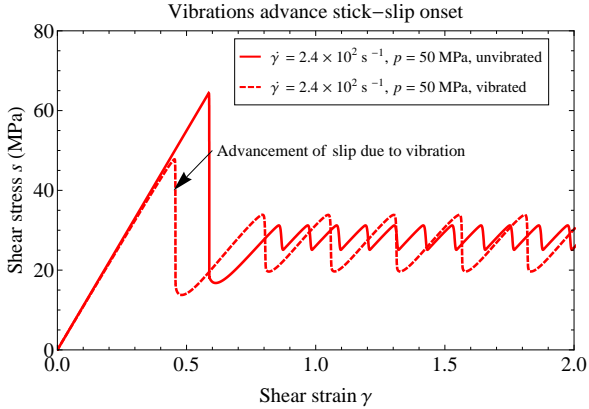}
\caption{\label{fig:tsplot}(Color online) Shear stress $s$ a s a function of accumulated shear strain $\gamma$, at applied shear rate $\dot{\gamma} = 2.4 \times 10^2$ s$^{-1}$, with and without external vibrations at fixed intensity. The onset of slip occurs earlier in the presence of vibrations, as shown by the location of the first stress drop on the dashed curve for vibrated sand. This observation is reminiscent of of the triggering of an earthquake by seismic waves arriving from elsewhere.} 
\end{figure}

Figure~\ref{fig:tsplot} shows that vibrations trigger the onset of slip, in the sense that vibrations clock-advance the first slip event at a reduced shear stress following the initial loading. This provides a possible explanation for dynamic earthquake triggering, the observation that seismic waves radiated by an earthquake can trigger catastrophic events elsewhere~\cite{griffa_2011,ferdowsi_2014b,marsan_2008}. The triggering mechanism is associated with weakening the threshold for slip events, rather than increasing the load.

\begin{figure}[ht]
\centering 
\includegraphics[width=.45\textwidth]{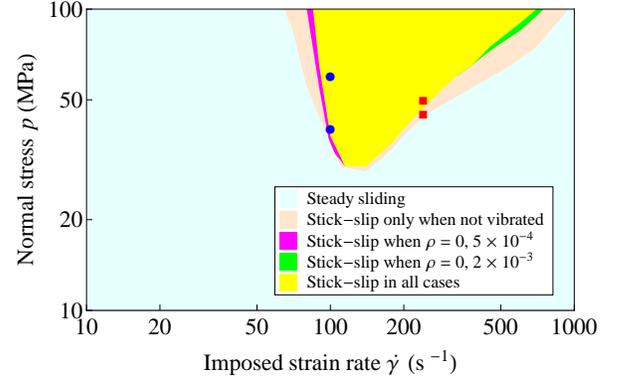}
\caption{\label{fig:phase}(Color online) Phase diagram showing the parameter space where stick-slip occurs, in the absence $(\rho = 0)$, and at two different, fixed vibration intensities $(\rho = 5 \times 10^{-4})$, $2 \times 10^{-3}$. Stick-slip operates only at imposed shear rates above $\dot{\gamma} \sim 80$ s$^{-1}$ and pressures above $p \sim 30$ MPa. Vibrations suppress stick-slip for a small range of pressure at all shear rates for which stick-slip is possible (shaded in pink), but amplify the stick-slip amplitude for a wide range of shear rates and pressure (shaded in yellow). There is a narrow range of shear rates and pressure (shaded in magenta and green) for which stick-slip is possible in the absence of vibrations, and for only one, but not both, of the two vibration intensities $\rho$. The blue circles and red squares denote to the shear rate and pressure values corresponding to the stress-strain curves in Figs.~\ref{fig:nofrplot} and~\ref{fig:zsplot}.} 
\end{figure}

Finally, we show in Fig.~\ref{fig:phase} our theoretical predictions regarding the pressure and shear rate regimes at which stick-slip operates. In addition to vibrations at intensity $\rho = 5 \times 10^{-4}$ as well as the unvibrated case studied earlier in the paper, we briefly examine here the effect of a higher vibration intensity $\rho = 2 \times 10^{-3}$ on stick-slip behavior. (This corresponds to doubling the vibration frequency, or quadrupling the amplitude. We set $\tilde{\chi}(\rho = 2 \times 10^{-3}) = 0.16$ to reflect the increased degree of compaction due to a higher vibration intensity.) We show that changing the vibration intensity $\rho$ shifts the boundary of the parameter regime for which vibrations suppress stick-slip that would otherwise have been possible. For our system of angular sand particles, there is a threshold normal stress $p_{\text{min}} \approx 30$ MPa and threshold shear rate $\dot{\gamma}_{\text{min}} \approx 80$ s$^{-1}$ below which only steady sliding is possible. Note from Eq.~\eqref{eq:sdot} that the ``effective spring stiffness'' $G/p$ is inversely proportional to the pressure, so the existence of a lower bound for the pressure corroborates with earlier results~\cite{daub_2009} that indicate the necessity of a low spring stiffness for stick-slip instabilities. For each pressure value $p$ above the threshold $p_{\text{min}}$, stick-slip can occur only for a range of intermediate shear rates. This corroborates with Fig.~\ref{fig:xvplot} which suggests that instabilities are possible only for a range of imposed strain rates.

Vibrations suppress stick-slip at the lower pressures at shear rates for which stick-slip would have been possible. Of interest is the narrow range of shear rates and pressures, shaded in green in Fig.~\ref{fig:phase}, for which stick-slip is possible only if the angular sand particles are unvibrated, or vibrated at the higher intensity $\rho = 2 \times 10^{-3}$, but that the lower-intensity vibration at $\rho = 5 \times 10^{-4}$ would suppress stick-slip behavior. The notion that some vibration amplitudes and frequencies can suppress stick-slip is corroborated by experimental observations~\cite{marone_private}, and may be of practical importance to control engineering. At high enough shear rates, shear-induced dilatation described by $\hat{\chi}(q)$ dominates over the effect of interparticle friction $\xi$ in Eq.~\eqref{eq:xdot}. As such, stick-slip becomes possible only at very high pressures which, in addition to reducing the effective stiffness, also decreases the inertial time scale, and hence the dimensionless plastic strain rate $q = \tau \dot{\gamma}^{\text{pl}}$ as well as the amount of dilation described by $\hat{\chi}(q)$.

\section{Concluding remarks}\label{sec:6}

We have proposed a theoretical model for dense granular flow. The model captures experimentally observed phenomena such as stick-slip instabilities. Additionally, it captures the clock advancement of slip due to external vibrations, accounting for the triggering of instabilities. The key ingredients of the model include: (a) noise produced by friction $\xi$, Eq.~\eqref{eq:xi}, which increases the packing fraction when the granular material undergoes shear deformation; (b) a $\chi$-dependent yield stress parameter $\mu_0$, Eq.~\eqref{eq:mu_0}, which produces apparent rate-weakening behavior at intermediate strain rates; and (c) explicit account of work associated with external vibrations, Eq.~\eqref{eq:first_law}. The underlying framework of nonequilibrium thermodynamics provides a natural way to incorporate energy dissipation as well as the effects of interparticle friction and external acoustic vibrations. The compactivity, an effective-disorder temperature, plays a crucial role in the model. Its dynamics encapsulate how interparticle frictional dissipation and vibrations control the response of a granular medium to shear deformation.

We found that noise generated by interparticle frictional dissipation constitutes an essential ingredient for stick-slip instabilities to occur, and found that this mechanism is most prominent at intermediate shear rates ($\dot{\gamma} \sim 10^3$ s$^{-1}$ for angular quartz sand) and high enough pressure. Additionally, we investigated the role of vibrations in controlling stick-slip behavior; we found that vibrations trigger slip events, change the stick-slip amplitude and period over a range of pressure and shear rate values, and suppress stick-slip in other circumstances. Within our choice of material parameters, we have computed a stick-slip phase diagram that illustrates the conditions on the shear rate and normal stress that are necessary for stick-slip to occur. The fact that our model predicts experimentally observed phenomena, including auto-acoustic compaction~\cite{lieou_2014b} and stick-slip instabilities, lends credence to the applicability of nonequilibrium thermodynamics to driven granular media.

A physics-based model such as the one laid out by us in the paper offers a predictive description of phenomena that may occur at regimes not currently accessible by laboratory experiments. Our theoretical model offers a unified understanding of the underlying mechanisms for stick-slip failure, as well as amplification and suppression of catastrophic events. Based on physical principles associated with grain-scale processes and nonequilibrium thermodynamics, these results shed light on practical problems in granular physics in earth science, industry, and beyond. For example, the notion that one can control stick-slip behavior by external acoustic vibrations may have applications in the nondestructive evaluation of materials, the control of jamming and unjamming, damage, and wear in materials processing, as well as in other industrial applications.

In the context of granular earth materials, an important example of phenomena observed in nature and in simulations, and captured in the present STZ model, is the dynamic triggering of catastrophic events~\cite{ferdowsi_2014a,ferdowsi_2014b,griffa_2011,marsan_2008}, known to occur in the presence of a granular gouge layer. Specifically, we found that vibrations reduce the shear strength and clock-advance the first slip event upon loading. This corroborates with the observation that seismic waves emanating from one earthquake can trigger another elsewhere. Thus the STZ theory provides important insight into the physics of fault failure in the earth in the presence of seismic waves.

For much of this paper we have used one single vibration intensity $\rho$ corresponding to a single set of values of vibration amplitude and frequency. We see from Fig.~\ref{fig:phase} that a change in vibration intensity can control stick-slip behavior. To further characterize vibration-induced shear weakening and dynamic triggering, and to identify intervals of increased seismic hazard, it will be useful to investigate the effect of varying the noise strength $\rho$ as well as the vibration amplitude and frequency. Ferdowsi \textit{et al}.~\cite{ferdowsi_2014a,ferdowsi_2014b} have found in their simulations evidence of an amplitude threshold below which vibrations do not cause significant frictional weakening. It will be interesting to examine if the STZ theory can capture this observation. This is beyond the scope of the present paper, but will be the subject of a future investigation.

Our model results display evidence of preseismic slip, shown in Figs.~\ref{fig:nofrplot} and~\ref{fig:pssplot} above. That is, the stress-strain curve shows some degree of rounding during the stick phase, corresponding to a gradual increase in the slip rate or plastic strain rate $\dot{\gamma}^{\text{pl}}$, prior to the onset of slip events. Characterization of the stress state immediately prior to shear failure, as well as the degree to which preseismic slip hints at the arrival of large slip events, may have important implications on the evaluation of seismic hazards.

In this paper we assumed for simplicity that the material is spatially homogeneous; in effect, we have considered only what happens within the shear band. However, real granular materials are heterogeneous in all spatial directions, and the shear band thickness depends on a variety of factors such as boundary conditions, imposed shear rate, and aggregate shear modulus~\cite{daub_2009,lieou_2014a,manning_2007a,manning_2009}. It may be necessary to consider a spatially heterogeneous compactivity distribution in order to fully capture irregular stick-slip dynamics seen in experiments such as~\cite{johnson_2008}.

We also assumed here for simplicity that grains do not break apart, in order to focus on the most salient physical ingredients associated with stick-slip. A simple mean-field physical model for grain fragmentation was discussed in earlier work~\cite{lieou_2014a}. There, we showed that grain fragmentation occurs readily above the crushing strength of particles, which is of the order of dozens of MPa and determined partly by the grain size. Changes in particle size may have important implications on the frictional noise term $\xi$, the shear rate regime in which instabilities may occur, and the shear strength of the granular medium. In \cite{lieou_2014a} we argued that grain fragmentation reduces the shear strength in a rate-strengthening granular material. One could also easily envision the physical picture of small grains produced by fragmentation getting trapped between large grains, causing compaction. These aspects are beyond the scope of the present paper, but could have important implications on the stick-slip amplitude and frequency~\cite{mair_2008}. The effect of grain fragmentation should be addressed in detail if one wants to examine the full range of possible physical behaviors in crushed rock particles.

Future studies will account for the roles of inertial effects, boundary conditions, the physics at the grain contacts, and grain fragmentation. These are in addition to addressing hysteretic phenomena and long-time relaxation upon ceasing acoustic vibrations or in unsheared granular media~\cite{griffa_2011,ferdowsi_2014a,ferdowsi_2014b}, as well as an extensive characterization of stick-slip statistics such as dependence of the magnitude of stress drop and inter-event period on the vibration intensity $\rho$ and other control parameters. While our results provide crucial input for refining the constitutive description of dense granular flow, we call for extensive experimental studies to validate our predictions and further constrain material parameters.

\section*{Acknowledgments}

We thank Paul Johnson for stimulating discussions. This work was supported by NSF Grant No. DMR0606092 and EAR-1345108, and the NSF/USGS Southern California Earthquake Center, funded by NSF Cooperative Agreement EAR-0529922 and USGS Cooperative Agreement 07HQAG0008, and the David and Lucile Packard Foundation. Additionally, JSL was supported in part by the U.S. Department of Energy, Office of Basic Energy Sciences, Materials Science and Engineering Division, DE-AC05-00OR-22725, through a subcontract from Oak Ridge National Laboratory.


\begin{thebibliography}{5}

\bibitem{johnson_2008}P. A. Johnson, H. Savage, M. Knuth, J. Gomberg, and C. Marone, Nature \textbf{451}, 57 (2008).

\bibitem{anthony_2005}J. L. Anthony and C. Marone, J. Geophys. Res. \textbf{110}, B08409 (2005).

\bibitem{mair_2002}K. Mair, K. M. Frye, and C. Marone, J. Geophys. Res. \textbf{107}, 2219 (2002).

\bibitem{han_2011}R. Han \textit{et al}., Geology \textbf{39}, 599 (2011).

\bibitem{yamashita_2014}F. Yamashita, E. Fukuyama, and K. Mizoguchi, Geophys. Res. Lett \textbf{41}, 341 (2014).

\bibitem{sone_2009}H. Sone and T. Shimamoto, Nature Geoscience \textbf{2}, 705 (2009).

\bibitem{daniels_2014}K. E. Daniels, C. Bauer, and T. Shinbrot, Graular Matter \textbf{16}, 217 (2014).

\bibitem{hayman_2011}N. W. Hayman, L. Ducloue, K. L. Foco, and K. E. Daniels, Pure. Appl. Geophys. \textbf{168}, 2239 (2011).

\bibitem{marone_private}C. Marone, private communication.

\bibitem{ferdowsi_2014a}B. Ferdowsi \textit{et al}., Phys. Rev. E \textbf{89}, 042204 (2014).

\bibitem{ferdowsi_2014b}B. Ferdowsi \textit{et al}., Acta Mech. \textbf{225}, 2227 (2014).

\bibitem{vanderelst_2010}N. J. van der Elst and E. E. Brodsky, J. Geophys. Res. \textbf{115}, B07311 (2010).

\bibitem{griffa_2013}M. Griffa \textit{et al}., Phys. Rev. E \textbf{87}, 012205 (2013).

\bibitem{daniels_2005}K. E. Daniels and R. P. Behringer, Phys. Rev. Lett. \textbf{94}, 168001 (2005).

\bibitem{daniels_2006}K. E. Daniels and R. P. Behringer, J. Stat. Mech. \textbf{2006}, P07018 (2006).

\bibitem{merrow_2000}E. W. Merrow, Chem. Innov. \textbf{30}, 34 (2000).

\bibitem{leuenberger_2005}H. Leuenberger and M. Lanz, Adv. Powder Tech. \textbf{16}, 3 (2005).

\bibitem{roberts_2002}A. W. Roberts and C. M. Wensrich, Chem. Engr. Sci. \textbf{57}, 295 (2002).

\bibitem{griffa_2011}M. Griffa \textit{et al}., Euro. Phys. Lett. \textbf{96}, 14001 (2001).

\bibitem{mair_2008}K. Mair and S. Abe, Earth Planet. Sci. Lett. \textbf{274}, 72 (2008). 

\bibitem{daub_2011}E. G. Daub, D. R. Shelly, R. A. Guyer, and P. A. Johnson, Geophys. Res. Lett. \textbf{38}, L10301 (2011).

\bibitem{edwards_1989a}S. F. Edwards and R. B. S. Oakeshott, Physica A \textbf{157}, 1080 (1989).

\bibitem{edwards_1989b}A. Mehta and S. F. Edwards, Physica A \textbf{157}, 1091 (1989).

\bibitem{edwards_1989c}S. F. Edwards and R. B. S. Oakeshott, Physica D \textbf{38}, 88 (1989).

\bibitem{edwards_1990a}S. F. Edwards, Rheol. Acta \textbf{29}, 493 (1990).

\bibitem{edwards_1990b}S. F. Edwards, J. Phys. Condensed Matter \textbf{2}, SA63 (1990).

\bibitem{haxton_2012}T. K. Haxton, Phys. Rev. E \textbf{85}, 011503 (2012).

\bibitem{lieou_2012}C. K. C. Lieou and J. S. Langer, Phys. Rev. E \textbf{85}, 061308 (2012).

\bibitem{lieou_2014a}C. K. C. Lieou, A. E. Elbanna, and J. M. Carlson, Phys. Rev. E \textbf{89}, 022203 (2014).

\bibitem{lieou_2014b}C. K. C. Lieou, A. E. Elbanna, J. S. Langer, and J. M. Carlson, Phys. Rev. E \textbf{90}, 032204 (2014).

\bibitem{vanderelst_2012}N. J. van der Elst, E. E. Brodsky, P.-Y. Le Bas, and P. A. Johnson, J. Geophys. Res. \textbf{117}, B09314 (2012).

\bibitem{dijksman_2011}J. A. Dijksman, G. H. Wortel, L. T. H. van Dellen, O. Dauchot, and M. van Hecke, Phys. Rev. Lett. \textbf{107}, 108303 (2011).

\bibitem{falk_1998}M. L. Falk and J. S. Langer, Phys. Rev. E \textbf{57}, 7192 (1998).

\bibitem{langer_2011}M. L. Falk and J. S. Langer, Ann. Rev. Cond. Matt. Phys. \textbf{2}, 353 (2011).

\bibitem{coleman_1963}B. D. Coleman and W. Noll, Arch. Ration. Mech. Anal. \textbf{13}, 167 (1963).

\bibitem{langer_2009a}E. Bouchbinder and J. S. Langer, Phys. Rev. E \textbf{80}, 031131 (2009).

\bibitem{langer_2009b}E. Bouchbinder and J. S. Langer, Phys. Rev. E \textbf{80}, 031132 (2009).

\bibitem{langer_2009c}E. Bouchbinder and J. S. Langer, Phys. Rev. E \textbf{80}, 031133 (2009).

\bibitem{nowak_1998}E. R. Nowak, J. B. Knight, E. Ben-Naim, H. M. Jaeger, and S. R. Nagel, Phys. Rev. E \textbf{57}, 1971 (1998).

\bibitem{knight_1995}J. B. Knight, C. G. Fandrich, C. N. Lau, H. M. Jaeger, and S. R. Nagel, Phys. Rev. E \textbf{51}, 3957 (1995).

\bibitem{edwards_1998}S. F. Edwards and D. V. Grinev, Phys. Rev. E \textbf{58}, 4758 (1998).

\bibitem{haxton_2007}T. K. Haxton and A. J. Liu, Phys. Rev. Lett. \textbf{99}, 195701 (2007).

\bibitem{manning_2007b}J. S. Langer and M. L. Manning, Phys. Rev. E \textbf{76}, 056107 (2007).

\bibitem{daub_2009}E. G. Daub and J. M. Carlson, Phys. Rev. E \textbf{80}, 066113 (2009).

\bibitem{daub_2010}E. G. Daub and J. M. Carlson, Ann. Rev. Cond. Matt. Phys. \textbf{1}, 397 (2010).

\bibitem{elbanna_2014}A. E. Elbanna and J. M. Carlson, J. Geophys. Res. Solid Earth \textbf{119}, 4841 (2014).

\bibitem{jop_2006}P. Jop, Y. Forterre, and O. Pouliquen, Nature (London) \textbf{441}, 727 (2006).

\bibitem{pechenik_2003}J. S. Langer and L. Pechenik, Phys. Rev. E \textbf{68}, 061507 (2003).

\bibitem{marsan_2008}D. Marsan and O. Lengeline, Science \textbf{319}, 1076 (2008).

\bibitem{manning_2007a}M. L. Manning, J. S. Langer, and J. M. Carlson, Phys. Rev. E \textbf{76}, 056106 (2007).

\bibitem{manning_2009}M. L. Manning, E. G. Daub, J. S. Langer, and J. M. Carlson, Phys. Rev. E \textbf{79}, 016110 (2009).





























\end{thebibliography}
\end{document}